\documentclass{NSF}
\usepackage{color, amssymb,amsmath,bm,setspace, enumitem, hyperref, subcaption, rotating}

\usepackage{caption}
\usepackage{subcaption}
\usepackage{multirow}
\usepackage{hhline}

\usepackage{booktabs}
\usepackage{dcolumn}
\usepackage{array}

\usepackage{url,times}
\usepackage{tabularx}
\usepackage{arydshln}
\usepackage{textcomp}
\usepackage{setspace}
\usepackage{enumitem}
\usepackage{siunitx}
\usepackage{hyperref}

\usepackage{booktabs,subcaption,amsfonts,dcolumn}

\usepackage{mathtools}
\usepackage[ruled]{algorithm2e}

\usepackage{setspace}

\begin{document}

\begin{center}
{\Large \textbf{Neural Speech and Audio Coding}}\\
Minje Kim\footnote{This work was supported in part by Electronics and Telecommunications Research Institute (ETRI) grant funded by the Korean government (24ZC1100;
``The research of the basic media contents technologies").\\\\
\centerline{ \textit{Full Citation: Minje Kim and Jan Skoglund,}}\\
\centerline{ \textit{``Neural Speech and Audio Coding: Modern AI technology meets traditional codecs,"}}\\
\centerline{ \textit{in IEEE Signal Processing Magazine, vol. 41, no. 6, pp. 85-93, Nov. 2024,}}\\
\centerline{\textit{doi: 10.1109/MSP.2024.3444318.}}
}, 
University of Illinois at Urbana-Champaign (\url{minje@illinois.edu})\\
Jan Skoglund, Google LLC (\url{jks@google.com})\\
\end{center}

\vspace{0.1in}

\centerline{
\begin{minipage}{13cm}
\textbf{Abstract}: This paper explores the integration of model-based and data-driven approaches within the realm of neural speech and audio coding systems. It highlights the challenges posed by the subjective evaluation processes of speech and audio codecs and discusses the limitations of purely data-driven approaches, which often require inefficiently large architectures to match the performance of model-based methods. The study presents hybrid systems as a viable solution, offering significant improvements to the performance of conventional codecs through meticulously chosen design enhancements. Specifically, it introduces a neural network-based signal enhancer designed to post-process existing codecs' output, along with the autoencoder-based end-to-end models and LPCNet--hybrid systems that combine linear predictive coding (LPC) with neural networks. Furthermore, the paper delves into predictive models operating within custom feature spaces (TF-Codec) or predefined transform domains (MDCTNet) and examines the use of psychoacoustically calibrated loss functions to train end-to-end neural audio codecs. Through these investigations, the paper demonstrates the potential of hybrid systems to advance the field of speech and audio coding by bridging the gap between traditional model-based approaches and modern data-driven techniques.    
\end{minipage}}

\section{Introduction}
\begin{figure}[h]
    \centering
    \includegraphics[width=.5\columnwidth]{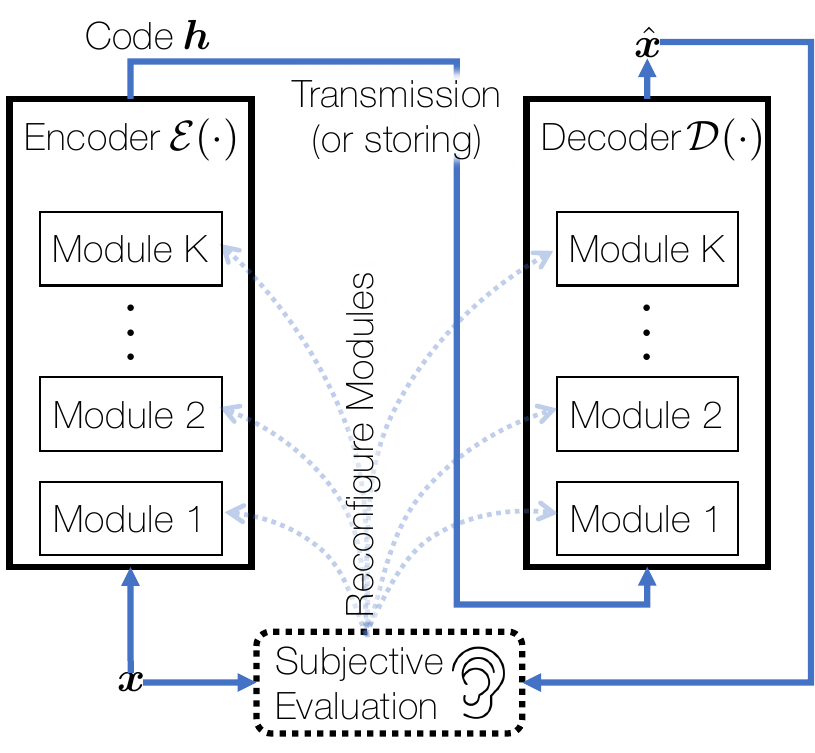}
    \caption{An ordinary development process of a model-based, legacy codec. The encoder and decoder are represented as functions, $\mathcal{E}(\cdot)$ and $\mathcal{D}(\cdot)$, respectively.}
    \label{fig:legacy-codec}
\end{figure}

Traditional speech and audio coding is a well-established technology, where various model-based approaches have been effective in compressing raw audio signals into compact bitstrings (encoding) and then restoring them to their original signal domain (decoding). These models aim to maintain the original signal's quality, such as speech intelligibility or other perceptual sound qualities, which are often subjectively defined. Hence, developing such models typically involves multiple rounds of listening tests to precisely measure the codec's performance. Although these models are designed by domain experts based on their knowledge and experience, finalizing them still requires tuning their parameters through listening tests and manual adjustments. Figure \ref{fig:legacy-codec} illustrates the ordinary development process of model-based coding systems.

The main challenge in traditional codec development is that manual tuning of model parameters relies significantly on time-consuming and costly listening tests. Successful models have emerged from extensive manual efforts and decades of research, including psychoacoustic models for bit allocation in MPEG-2 Audio Layer III (a.k.a. MP3) \cite{mp3, PainterT2000ieeeproc}, spectral band replication \cite{sbr}, time-frequency transformation methods such as the modified discrete cosine transform (MDCT) \cite{aac-standard}, and speech generation models like linear predictive coding (LPC) \cite{amr-wb-standard}, to name a few. However, standard speech and audio codecs typically consist of multiple heterogeneous coding blocks that interact with one another, complicating the tuning task.

Meanwhile, data-driven approaches have also been successfully introduced to conventional codecs as their components. For example, unified speech and audio coding (USAC) \cite{usac1, usac2} employs a classification module to detect the transient events from the input signal and then use the appropriate coding module accordingly, i.e., by increasing the temporal resolution of the TF transform in the transient period, and vice versa. Enhanced voice services (EVS), the latest 3GPP standard for voice communication, comes with the voice activity detection (VAD) feature, or more precisely, signal activity detection (SAD) \cite{evs-vad}. Due to its discriminative nature, which specifies whether each 20 ms frame contains a meaningful signal, it can be essentially considered a binary classification module, too. Finally, Opus, a royalty-free and open-source audio codec, also utilizes a small neural network-based classifier for VAD and music/speech classification \cite{opus}. Training those classification models in a completely data-driven way is not entirely straightforward due to the fact that the ground-truth labels of these sound events, and consequently, the classification performance, should be defined to improve the perceptual quality of the decoded signals. The intricacy hinders the researchers from investigating structural variations of the coding system, compared to a typical supervised learning setup where pre-defined classes manually label training samples.

\begin{figure}[h]
    \centering
    \includegraphics[width=.5\columnwidth]{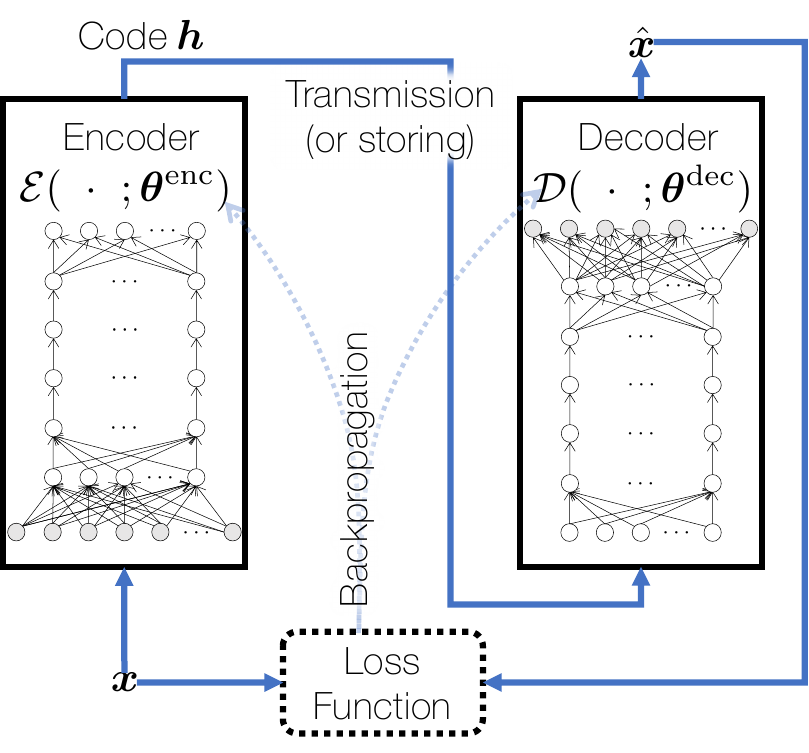}
    \caption{An end-to-end neural codec trained in a data-driven way. The encoder and decoder are now a parametric function, parameterized by $\bm\theta^{\text{enc}}$ and  $\bm\theta^{\text{dec}}$, respectively. Loss functions are employed to compute the reconstruction error, which is used to update the model parameters via gradient descent methods, e.g., backpropagation.}
    \label{fig:e2e-codec}
\end{figure}

Recently, data-driven methods for coding have gained intensive attention. In these neural speech and audio coding (NSAC) systems, a simple approach is to postulate an autoencoding task that reconstructs the input signal as exactly as possible in its output. In the meantime, to achieve the signal compression goal in the autoencoding pipeline, it is typical to decompose the system into two parts: an encoder module that produces a latent representation, followed by a decoder that recovers raw signal from the intermediate variable (Figure \ref{fig:e2e-codec}). Hence, it is critical to impose additional constraints on the encoder output so that it works as the \textit{code} for transmission. For example, the code vector is supposed to be low in dimensionality and robust to quantization for the maximum coding gain. Such autoencoders can be trained in an end-to-end fashion with an adequately defined reconstruction loss as well as an additional entropy control mechanism. Kankanahali proposed a neural speech coding model \cite{KankanahalliS2018icassp} that consists of fully convolutional encoder and decoder modules, whose bottleneck feature map is discretized via the soft-to-hard quantization method \cite{AgustssonE2017softmax}. Although this seminal end-to-end setup was successful to some degree, it fell short of competing with the legacy speech codec, e.g., adaptive multi-rate wideband (AMR-WB) \cite{amr-wb-standard}. More recently, such end-to-end autoencoder-based codecs have achieved higher coding gain, e.g., Soundstream \cite{Zeghidour2021soundstream}, EnCodec \cite{DefossezA2023encodec}, and Descript Audio Codec (DAC) \cite{KumarR2023dac}, where residual vector quantization and adversarial loss functions play big roles.  Another line of seminal work chose a powerful decoder architecture. A highly effective generative model, WaveNet \cite{OordA2016wavenet}, was employed as the decoder, which is trained to re-synthesize the speech signal by being indirectly conditioned on the very low-bitrate deterministic codes defined in the Codec 2 system \cite{codec2} as shown in \cite{KleijnW2018wavenet} or codes learned from vector-quantized variational autoencoder (VQ-VAE) as in \cite{GarbaceaC2019vqvae}. While these systems achieved phenomenal perceptual performance, they came at the cost of WaveNet's high computational complexity.

In this paper, we review the recent literature and introduce efforts that merge the model-based and data-driven approaches to improving speech and audio codecs. In addition to the prior review article that focused more on speech coding \cite{OShaughnessyD2023speechcoding}, we provide novel perspectives to the recent NSAC literature by encompassing various important concepts, such as generative, predictive, and psychoacoustic models, in the context of reviewing data- and model-driven approaches as well as their harmonization.  We begin with a general introduction to speech and audio codecs, including how they are related to some important application scenarios and what kind of requirements are commonly imposed on codecs. Based on this, we will situate the NSAC systems in the context of coding technologies. Then, we move on to a straightforward serialization of the traditional codec and a signal enhancement model that is trained to improve coding artifacts. Subsequently, we introduce exemplar codecs that achieved a more structural integration of the two different methods: one that harmonizes the recurrent neural networks' prediction power into the LPC process and another trend where such learning-based prediction improves transform-based audio codecs. Finally, we see empirical evidence that the well-established psychoacoustics model can inform the learning process by manipulating the loss function.

\section{Potentials and Limitations of NSAC}\label{sec:potential}

Modern speech and audio codecs can be grouped into two categories based on their application. First, codecs are heavily used in communication scenarios where multiple end users are involved. The predominant content, in this case, is spoken language. Hence, codecs are required to process speech in real-time or with a delay low enough for people not to recognize. Digital voice communication applications that use voice over Internet Protocol (VoIP) are representative examples. Since voice communication can happen between mobile (i.e., resource-constrained) devices nowadays, it is also very important for codecs to operate with minimal computational and spatial complexity. In addition, there is an increasing need to handle mixed contents in communication scenarios, such as non-speech signals mixed with users' utterances, necessitating communication codecs to be robust to various contents in adversarial environments. Finally, depending on the bandwidth of the communication channel, this type of codec may be required to operate in low bitrates, e.g., lower than 10 kbps. 

Second, codecs can be used in a more uni-directional application scenario, such as for media streaming, digital broadcasting, storing music signals, etc. In these uni-directional applications, the user on the decoder side tends to be less sensitive to the delay. Instead, listeners are often more sensitive to the subtle discrepancies and artifacts that the codecs generate, e.g., in music listening. Consequently, these media codecs are designed to provide a high-fidelity reconstruction for various input audio signals, including speech, music, mixed, and multichannel contents.

Although NSAC systems are evolving in the direction that tries to meet the requirements mentioned above, they possess inherent characteristics that come from their data-driven nature. One of the most distinctive factors is the high computational complexity, which ranges from 100G floating-point operations per second (FLOPS) in a large autoregressive model, e.g., a WaveNet decoder \cite{KleijnW2018wavenet}, to a relatively efficient LPCNet decoder, which still requires 3 GFLOPS in its original version \cite{ValinJ2019lpcnetcoding}. Meanwhile, one of the standard speech codecs, AMR-WB, can decode with only 7.8 weighted million operations per second (WMOPS) \cite{BessetteB2002amrwb}. Likewise, although a direct and rigorous comparison is impossible, neural codecs are multiple orders of magnitudes more complex than traditional codecs. While it is expected that more advanced hardware architecture for neural network inference could make neural codecs more affordable on devices, it is clear that reducing computational complexity is a common goal in designing neural codecs for on-device processing. In addition, the high computational complexity hinders real-time processing as the system consumes much more resources to process every frame in time.

On the other hand, some neural speech codecs have shown sensible improvement in compression ratio while keeping the same level of speech quality with traditional codecs, i.e., from ``Good" \cite{KleijnW2018wavenet, ValinJ2019lpcnetcoding} to ``Excellent" \cite{GarbaceaC2019vqvae} in multiple stimuli with hidden reference and anchor (MUSHRA) \cite{mushra} at the cost of higher complexity. Neural speech codecs' very low bitrates can also improve their robustness to packet loss because the low bitrates allow for redundantly many packets for communication \cite{ValinJM2023icassp}. Another potential convenience in developing neural codecs is that it is relatively more straightforward to develop a universal codec that works for both speech and non-speech audio signals by training the model with various types of audio signals in a data-driven fashion. However, so far, the literature has shown that dedicated neural speech codecs can achieve better speech reconstruction than attempting to handle general audio signals. Indeed, general-purpose high-fidelity audio coding still remains a challenge for neural audio codecs despite substantial potentials shown in some neural audio codecs in low bitrates ($<$ 12 kbps) \cite{Zeghidour2021soundstream, DefossezA2023encodec, KumarR2023dac}. Note that the subjective listening test results reported in various NSAC papers are not directly comparable as they were done with different combinations of coding systems, adding more subjectivity to the individual results. Instead, in this paper we discuss their general trend. In addition, we opt to use the term NSAC to encompass both neural speech and audio codecs, because they can be potentially extended to cover both types of signals in the future.

\section{Data-Driven Approaches to Removing Coding Artifacts}\label{sec:se}

Depending on the characteristics of the underlying model, a codec can produce unique coding artifacts, which can lower the perceptual quality of the decoded signal. In theory, we can postulate a general-purpose signal enhancement system, trained to reduce various types of coding artifacts by mapping decoded signals to their original inputs. However, in practice, learning such a mapping function through a data-driven approach by focusing on a specific type of codec and the signals it processes is more feasible. Let $\mathcal{F}(\cdot)$ represent a legacy codec, which can be decomposed into the encoder $\mathcal{E}(\cdot)$ and decoder $\mathcal{D}(\cdot)$ modules:
\begin{equation}\label{eq:ae}
    \bm{x}\approx\hat{\bm{x}}\leftarrow \mathcal{F}(\bm x)=\mathcal{D}\circ\mathcal{E}(\bm x).
\end{equation}

With this framework, we can propose an additional parametric model trained to map the decoded signal $\hat{\bm{x}}$ back to the original, unprocessed input $\bm x$, 
\begin{equation}
    \bm x \approx \tilde{\bm x}\leftarrow\mathcal{G}(\hat{\bm x}; \bm\theta), 
\end{equation}
where the training process, e.g., a variation of gradient descent, updates the model parameter $\bm \theta$ to the direction that can minimize $\mathcal{L}(\bm x||\tilde{\bm x})$, which is a pre-defined metric that measures the difference between the two signals. In other words, with the additional denoising process introduced by $\mathcal{G}(\cdot)$, we hope the quality of the decoded signal improves in terms of the loss function $\mathcal{L}(\cdot)$:
\begin{equation}
    \mathcal{L}(\bm x||\tilde{\bm x}) < \mathcal{L}(\bm x||\hat{\bm x}).
\end{equation}

This pipeline is convenient to implement because the only learnable module, $\mathcal{G}(\cdot; \bm\theta)$, can be concatenated with any existing codec after it has been trained to enhance $\hat{\bm{x}}$ into $\tilde{\bm{x}}$. It is because the underlying codec, $\mathcal{F}(\cdot)$, is presumed to be already fully configured or \textit{frozen} from the perspective of the learning algorithm. 

Moreover, this post-processing approach does not increase bitrate. Typically, signals in communication are transmitted in encoded data, i.e., $\bm h \leftarrow \mathcal{E}(\bm x)$, with the decoder located on the receiver side. Therefore, any operations performed after decoding, such as $\mathcal{G}(\hat{\bm x}; \bm\theta)$, do not impact the bitrate or alter the behavior of the encoder, rendering the post-processing module a \textit{bitrate-free} signal enhancer.

The drawback of this approach is that the codec and enhancement module are segregated. The ability to utilize any existing codec also implies that the codec's inherent characteristics remain unchanged and do not directly benefit from the data-driven approach. Consequently, the enhancement module $\mathcal{G}(\cdot;\bm\theta)$ may struggle to eliminate codec-specific artifacts, necessitating substantial model capacity and increasing decoding complexity.

\subsection{Supervised Signal Enhancement Models for Post-Processing}
In \cite{ZahoZ2019cnn_enhance_coded_speech}, convolutional neural network (CNN) models were introduced to enhance coded speech, serving as an alternative to the $\mathcal{G}(\cdot; \bm\theta)$. The authors experimented with two distinct versions, one operating directly in the time domain in an end-to-end process, and another leveraging cepstrum features. In comparison to traditional post-filtering methods employed by G.711 \cite{postefilter}, their approach demonstrated noticeable improvements across various objective metrics and listening tests. This method closely parallels the supervised DNN-based speech enhancement problem, where the primary objective is to eliminate any undesired artifacts from real-world speech recordings. Specifically applied to coded speech, the model was trained to remove the coding artifacts. 

As for audio coding, both CNNs and recurrent neural networks (RNN) have been utilized to enhance the MP3-compressed signals \cite{DengJ2019lstm-mp3-restoration}. Among them, a long short-term memory (LSTM) network has been effectively used to predict signals in the time and frequency domains defined by the modified discrete cosine transform (MDCT), referred to as T- and F-LSTM, respectively. The TF-LSTM method improved the subjective quality in terms of mean opinion score (MOS), particularly when the post-processing is applied to the more demanding stereophonic signals at 96kbps or on the 64kbps mono signals.

\subsection{Generative Models as a Post-Processor to Enhance Coded Speech}

Since the coding artifacts primarily stem from information loss rather than additive noise, supervised learning-based approaches may face challenges in imputing missing values. Therefore, exploring a more generative approach for addressing the codec-specific enhancement issue is justifiable. 

Biswas and Jia proposed a generative adversarial network (GAN) \cite{GoodfellowI2014gan} to enhance the performance of the AAC codec \cite{aac-standard} at low bitrates, namely deep coded audio enhancer (DCAE) \cite{BiswasA2020gan-coding}. Utilizing the GAN formulation, the enhancement network functions as the \textit{generator}, trained to generate realistic examples from random noise $\bm n$ drawn from the standard normal distribution,
\begin{equation}\label{eq:gen}
    \tilde{\bm x}\leftarrow\mathcal{G}(\hat{\bm x}, \bm n; \bm\theta_\mathcal{G}), ~ \bm n \sim \mathcal{N}(\bm 0, \bm 1).
\end{equation}
Note that, the enhancement network $\mathcal{G}(\cdot)$  now acts as a generator, which takes both the decoded audio $\hat{\bm{x}}$ to be enhanced and random noise $\bm n$ to add stochasticity to the process. 

Furthermore, the generated examples are examined by a \textit{discriminator} $\mathcal{C}(\cdot; \bm\theta_\mathcal{C})$, which acts as a binary classifier trained to differentiate between synthetic and real-world examples. Consequently, the ultimate objective of GAN training is to reach a \textit{Nash equilibrium}, a state where the discriminator fails to tell the difference between the real and fake examples. To mitigate the AAC codec's artifacts, the original GAN formulation was modified in the following manner:
\begin{align}
    \label{eq:gen_loss}
    \mathcal{L}_\mathcal{G}(\bm\theta_\mathcal{G})&=\mathbb{E}_{{\bm n}\sim \mathcal{N}(\bm 0, \bm 1),~{\hat{\bm{ x}}\sim p_{\text{data }}(\bm{ x})}} \big[\big(\mathcal{C}(\tilde{\bm x}, \hat{\bm x}; \bm\theta_\mathcal{C})-1\big)^2\big]+\lambda\mathbb{E}_{{\bm{ x}\sim p_{\text{data }}(\bm{ x})}}||\tilde{\bm x} - \bm x||_1,\\
    \label{eq:dic_loss}
    \mathcal{L}_\mathcal{C}(\bm\theta_\mathcal{C})&=\mathbb{E}_{\bm{x},\hat{\bm{ x}}\sim p_{\text{data }}(\bm{x})}\big[\big(\mathcal{C}({\bm x}, \hat{\bm x}; \bm\theta_\mathcal{C})-1\big)^2\big]+\mathbb{E}_{{\bm n}\sim \mathcal{N}(\bm 0, \bm 1),~{\hat{\bm{ x}}\sim p_{\text{data }}(\bm{ x})}}\big[\mathcal{C}(\tilde{\bm x}, \hat{\bm x}; \bm\theta_\mathcal{C})^2\big].
\end{align}

Eq. \eqref{eq:gen_loss} defines the loss $\mathcal{L}_\mathcal{G}$ as a function of the generator parameters $\bm{\theta}_\mathcal{G}$.
Beyond the standard reconstruction loss $||\tilde{\bm x} - \bm x||_1$ term, the first loss term also encourages the generator output $\tilde{\bm x}$ to be classified as 1, the ``real" examples category, by deceiving the discriminator. Note that the expectation spans two distributions, $\mathcal{N}(\bm 0; \bm 1)$, responsible for generating the random noise $\bm n$ that introduces stochasticity into the generator (eq. \eqref{eq:gen}), and the sample distribution of the training set $p_\text{data}(\bm x)$, whose sample deterministically goes through the coding process and defines the decoded signal $\hat{\bm x}$, too. In doing so, the fake example $\tilde{\bm x}$ is appended by the decoded signal $\hat{\bm x}$ to give more context to the classifier, i.e., $\mathcal{C}((\tilde{\bm x}, \hat{\bm x})$. Likewise, aside from the typical reconstruction loss $||\tilde{\bm x} - \bm x||_1$, within the GAN context, the generator's performance is evaluated only by the fake example pairs, $(\tilde{\bm x}, \hat{\bm x})$. 

On the contrary, the discriminator $\mathcal{C}(\cdot)$, as a classifier, also requires exposure to real example pairs, $({\bm x}, \hat{\bm x})$. In this case, since the examples are real, they must also be assigned label 1. The discriminator loss $\mathcal{L}_\mathcal{C}(\bm\theta_\mathcal{C})$ is defined over the discriminator parameters $\bm\theta_\mathcal{C}$, aiming to minimize the disparity between the predicted class of the real example pair $({\bm x}, \hat{\bm x})$ and label 1, while pushing the prediction of the fake example pair $(\tilde{\bm x}, \hat{\bm x})$ to be near zero. Note that the second term of the discriminator loss $\mathcal{L}_\mathcal{C}(\bm\theta_\mathcal{C})$ and the generator loss $\mathcal{L}_\mathcal{G}(\bm\theta_\mathcal{G})$ are in conflict, contributing to the instability of GAN training. 

The GAN-based method significantly enhanced the AAC codec's performance at 24 and 32 kbps, achieving an impressive gain of 10 to 14 points on average in MUSHRA tests for both speech and applause signals. While the baseline AAC codec is designed for general-purpose audio coding, the GAN training was specifically conducted on speech dataset or applaud samples, each treated independently. 

PostGAN introduces a more sophisticated approach, featuring advanced functionalities, such as subband processing and online processing with a minimal algorithmic delay of 10 ms \cite{KorseS2022postgan}. Originating from the low-delay Bluetooth codec, LC3 \cite{LC3}, PostGAN has demonstrated superior capabilities in enhancing the speech quality of decoded signals. 

\section{Learning to Predict Speech Signals}

The supervised signal enhancement models as well as their GAN variations introduced a sensible quality improvement to the traditional coding pipeline. Although the detached nature of the post-processor and the codec is convenient, the pipeline leaves room for structural innovation.

We start with the modeling of speech signals. It has long been known in the literature that a source-filter model can effectively explain the speech generation process \cite{FantG1971source-filter}. In this model, the source signal produced by the glottal vibration is filtered by the vocal tract to produce the formant effects. Traditional speech codecs have widely used the simple linear predictive coding (LPC) scheme to model the speech signal, by isolating the vocal tract's effects from the spectrum via a simple linear prediction model, 
\begin{equation}
    x_t \approx u_t\leftarrow\sum_{\tau=1}^p a_\tau x_{t-\tau},\quad e_t = x_t-u_t
\end{equation}
where $p$ is the order of LPC and $e_t$ stands for the residual between the sample $x_t$ and the prediction $u_t$, which could be a low-energy, quasi-periodic, and impulse-like signal if modeling is successfully done on the periodic component of speech, e.g., voiced areas. Once quantized, the LPC coefficients $a_\tau$ account for relatively low bitrates, e.g., 2.4 kbps via multistage vector quantization in AMR-WB, whose total bitrate varies from 6.6 to 23.85 kbps. Therefore, the codec's performance depends on how much coding gain it achieves in compressing the residual signal $e_t$. 

\subsection{End-to-End Codec for LPC Residual Coding}

\begin{figure}[h]
    \centering
    \includegraphics[width=0.6\columnwidth]{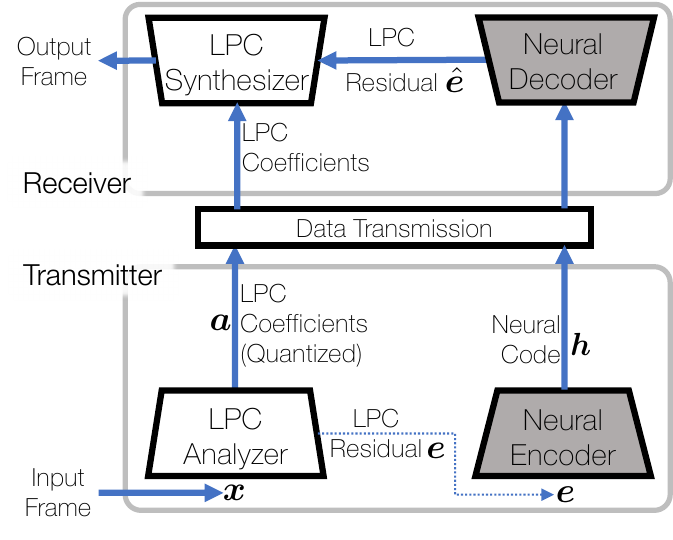}
    \caption{LPC as a pre-processor of an end-to-end coding system. }
    \label{fig:lpc+e2e}
\end{figure}

A straightforward way to combine neural coding and LPC is to replace the traditional speech coding module that operates in the LPC residual domain with an end-to-end autoencoder. Hence, eq. \eqref{eq:ae} can be redefined by taking the residual signal $\bm e$ as follows:
\begin{equation}
    \bm e \approx \hat{\bm e} \leftarrow \mathcal{F}(\bm e)=\mathcal{D}\circ\mathcal{E}(\bm e).
\end{equation}
In other words, if the autoencoding performance improves, a better residual reconstruction on the decoder side contributes to better LPC synthesis. Figure \ref{fig:lpc+e2e} depicts the general concept.  

In \cite{ZhenK2019interspeech}, this concept was empirically proved: instead of compressing the raw speech signal directly via a CNN-based autoencoder as in \cite{KankanahalliS2018icassp}, coding in the LPC residual domain can achieve a better perceptual quality at the same bitrate. In addition to the simple concatenation of LPC and CNN, the follow-up work investigated the dynamic bit allocation option between the LPC and the autoencoder modules depending on the type of the signal at the moment, increasing the coding gain even further \cite{ZhenK2022taslp}. In this line of work, cross-module residual learning (CMRL), LPC acts as the first module that explains the input signal, whose residual signal is modeled by the subsequent end-to-end autoencoders, one after another.  

\subsection{LPCNet}

\begin{figure}[h]
    \centering
    \includegraphics[width=0.5\columnwidth]{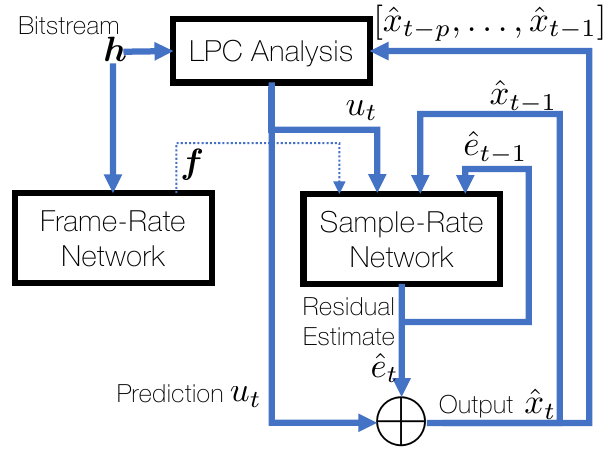}
    \caption{The LPCNet codec's synthesis process on the receiver side. This figure was redrawn based on the original one in \cite{ValinJ2019lpcnetcoding}.}
    \label{fig:lpcnet}
\end{figure}

A more sophisticated approach to combining the model-based and data-driven methods is LPCNet. It was originally proposed as a general-purpose vocoder, and then soon developed into a codec by being able to synthesize speech from a very low-bitrate (1.6 kbps) cepstrum-based code \cite{ValinJ2019lpcnetcoding}. The main idea of LPCNet-based vocoding on the receiver side is, first, to estimate the LPC coefficients $a_\tau$ from the quantized cepstrum code $\bm h$. Secondly, the transmitter does not send the residual signal $\bm e$, although it is essential for LPC synthesis. Instead, LPCNet employs an RNN-based data-driven module that predicts the missing residual signal from the other available information within the decoder, such as the past samples generated so far $[\hat{x}_{t-p}, \ldots, \hat{x}_{t-1}]$, LPC prediction $u_t$, and an auxiliary feature vector an auxiliary feature vector $\bm f$ induced from the code $\bm h$ (Figure \ref{fig:lpcnet} for an illustration):
\begin{align}
    \bm f &\leftarrow\mathcal{D}_\text{frame}(\bm h)\\
    u_t, a_1, \ldots, a_p &\leftarrow\mathcal{D}_\text{LPC}(\bm h, [\hat{x}_{t-p}, \ldots, \hat{x}_{t-1}])\\
    \hat{e}_t&\leftarrow\mathcal{D}_\text{sample}(\hat{e}_{t-1},\hat{x}_{t-1}, u_t, \bm f)\\
    \hat{x}_t&=\hat{e}_t+u_t,
\end{align}
where the \textit{frame-rate} network $\mathcal{D}_\text{frame}(\bm h)$ converts the code of the given frame into a feature vector $\bm f$. Meanwhile, $\mathcal{D}_\text{LPC}(\cdot)$ conducts the LPC coefficients estimation, and consequently, the prediction of the next sample $u_t$, from the code $\bm h$ and the past samples predicted by the LPCNet vocoder so far $[\hat{x}_{t-p}, \ldots, \hat{x}_{t-1}]$. Finally, the sample-rate network $\mathcal{D}_\text{sample}(\cdot)$ conducts the estimation of the residual signal $\hat{e}_t$ out of all available information, i.e., the previous estimation of the residual $\hat{e}_{t-1}$ and the reconstructed utterance sample $\hat{x}_{t-1}$, LPC prediction $u_t$, and the frame-rate feature $\bm f$.

LPCNet employs a WaveRNN \cite{KalchbrennerN2018wavernn} architecture based on the gated recurrent units (GRU) \cite{ChoK2014emnlp}. Trained to predict the residual sample, the transmitter is liberated from the burden of compressing the residual signal. Meanwhile, since the code is still based on the cepstrum and pitch information, the frame-level feature inferred from it further assists the synthesis process. Finally, LPCNet attains low computational complexity, suitable enough for real-time communication applications.

\section{Learning to Predict Speech and Audio Signals in the Feature Space}

A widely used principle in coding, as in LPC-based ones, is to use a predictive model, assuming that the distribution of the residual samples between the original and predicted signals is with lower entropy. Since entropy serves as the lower bound of the bitrate, residual coding is generally beneficial once the model's prediction is good enough.

\begin{figure}[h]
    \centering
    \includegraphics[width=0.55\linewidth]{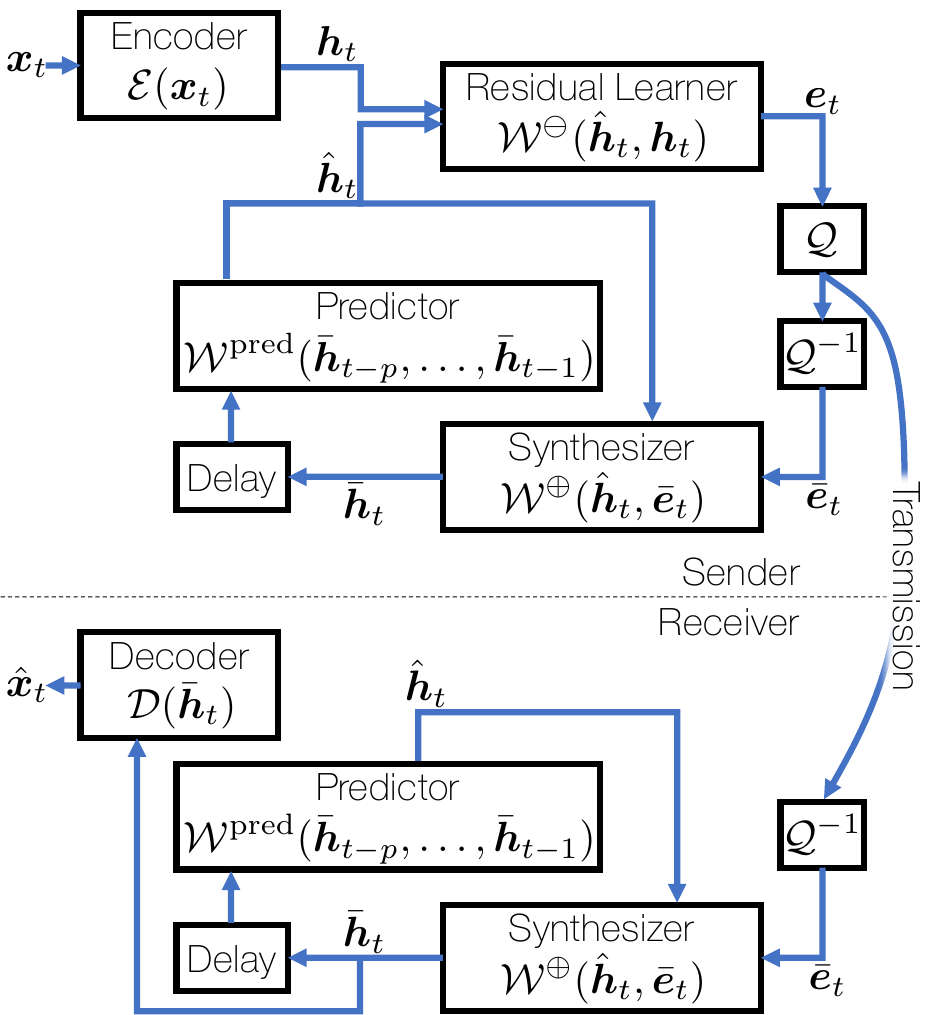}
    \caption{Residual coding in the feature space. $\bm h_t$ and $\bm e_t$ stand for the code and residual vectors at the $t$-th frame, while the $~\hat{}~$ and $~\bar{}~$ notations indicate that the variables are predicted and quantized versions, respectively.}
    \label{fig:prediction}
\end{figure}

Meanwhile, the basic idea behind an end-to-end neural codec, as described in Figure \ref{fig:e2e-codec}, is to convert the raw signal into the feature space where a code vector $\bm h$ can be quantized more effectively. Since quantization happens in the feature space in neural codecs, it is natural to employ residual coding in the feature space rather than in the raw signal domain.

Figure \ref{fig:prediction} illustrates the general residual coding concept implemented in the feature space. First, the raw samples at the $t$-th frame $\bm x_t$ is converted into the feature space, where quantization could be conducted if it were not for residual coding. The transformation can be done in a model-based approach, e.g., by using MDCT, but a data-driven method can also learn a custom feature space, e.g., by a CNN encoder $\mathcal{E}(\cdot)$. 

Next, on the sender side, the codec utilizes a predictor module $\mathcal{W}^\text{pred}(\cdot)$ to predict the current code vector $\bm h_t$ from its preceding feature vectors. In theory, $\mathcal{W}^\text{pred}(\cdot)$ can be trained using $p$ ground-truth features in the past, i.e., $[\bm h_{t-p}, \ldots, \bm h_{t-1}]$, but a more robust set up is to use a series of \textit{reconstructed} features $[\bar{\bm h}_{t-p}, \ldots, \bar{\bm h}_{t-1}]$ that are the reconstructed feature vectors that the receiver accumulates in its buffer to use as the \textit{past} examples for prediction. Hence, we write the prediction function, 
\begin{equation}\label{eq:prediction}
    \hat{\bm h}_t\leftarrow\mathcal{W}^\text{pred}(\bar{\bm h}_{t-p}, \ldots, \bar{\bm h}_{t-1}) 
\end{equation}
The reason why the predictor becomes more robust when it was trained to predict from the reconstructed features is that, in this way, the predictor is trained to encompass the potential reconstruction error in its prediction process, which the receiver has to deal with during the test-time inference. Note that the same predictor module is shared between the sender and receiver. 

The prediction vector $\hat{\bm h}_t$ is compared against the original feature $\bm h_t$, and then the residual signal $\bm e_t$ is induced. Here, the residual can be computed in a straightforward manner, i.e., $\bm e_t \leftarrow \bm h_t - \hat{\bm h}_t$, although a more involved \textit{residual learner} can be employed as follows:
\begin{equation}\label{eq:residual_learner}
    \bm e_t \leftarrow \mathcal{W}^\ominus(\hat{\bm h}_t, \bm h_t). 
\end{equation}

After quantization, the residual feature vector $\bm e_t$ is sent to the receiver for decoding. At the same time, the dequantized version $\bar{\bm e}_t$ is harmonized with the prediction $\hat{\bm h}_t$ back again to form the reconstruction $\bar{\bm h}_t$. Once again, the reconstruction can be as simple as a summation, i.e., $\bar{\bm h}_t\leftarrow \bar{\bm e}_t + \hat{\bm h}_t$, but it could also employ a neural network module that \textit{synthesizes} the reconstruction as follows:
\begin{equation}\label{eq:synthesizer}
    \bar{\bm h}_t\leftarrow \mathcal{W}^\oplus(\hat{\bm h}_t, \bar{\bm e}_t ). 
\end{equation}

The system can choose to utilize $p$ such reconstructed features as a sequential input to the predictor module $\mathcal{W}^\text{pred}(\bar{\bm h}_{t-p}, \ldots, \bar{\bm h}_{t-1})$, whose prediction $\hat{\bm h}_t$ is fed back to the pipeline as the input to the residual learner (eq. \eqref{eq:residual_learner}) and synthesizer (eq. \eqref{eq:synthesizer}).

The receiver operates similarly: (a) it reconstructs $p$ features using the same synthesizer as in eq. \eqref{eq:synthesizer} using the transmitted residual signal after dequantizing it, $\bar{\bm e}_{t}$, as well as the corresponding frame's prediction $\hat{\bm h}_{t}$ (b) the reconstructed features $\bar{\bm h}_t$ are accumulated for predicting the next frame's feature $\hat{\bm h}_t$ (c) finally, a decoder function converts the reconstructed feature back to the raw signal $\hat{\bm x}_t$.

Likewise, the merit of this approach is that the neural codec can additionally benefit from the temporal context of the signal, which can span longer than learning from the raw signal due to the frame-level prediction that decimates the original temporal resolution. As in other model-based approaches, once the prediction model is successful, the residual signal's energy reduces, which usually leads to a lower entropy. 

\subsection{Feature Prediction for the LPCNet Codec}

A GRU-based predictive model \cite{YangH2023lpcnet} introduced additional coding gain to the LPCNet codec, which corresponds to the prediction module in eq. \eqref{eq:prediction}. It aligns with the general concept shown in Figure \ref{fig:prediction}, except for its own specific configurations. First of all, since it employs a GRU model, which uses its hidden units $\bm y$ to summarize the past information, the prediction model does not need to maintain a buffer of past feature reconstructions. We can rewrite this GRU-based predictor as follows:
\begin{equation}
\hat{\bm h}_{t}, \bm y_{t}\leftarrow \mathcal{W}^\text{pred}(\bar{\bm h}_{t-1}, \bm y_{t-1}),
\end{equation}
where $\bm y_{t-1}$ essentially summarizes all previous features the GRU model has been exposed to.  

Another unique setup is that, instead of using a neural network encoder $\mathcal{E}(\cdot)$, it inherits the cepstrum-based code space that LPCNet uses. In addition, the residual signal and the reconstruction are conducted via the simple subtraction and addition operations, i.e., 
\begin{equation}
\bm e_t \leftarrow \bm h_t-\hat{\bm h}_t, \quad \bar{\bm h}_t\leftarrow \hat{\bm h}_t+\bar{\bm e}_t. 
\end{equation}

Finally, as for the decoder that recovers raw signals from the predicted feature $\hat{\bm h}_t$, the original LPCNet is directly used as a vocoder. The residual coding scheme introduced additional coding gain, e.g., about 5 points higher at a lower bitrate (1.47 kbps) than LPCNet's (1.6 kbps) in the MUSHRA test. 

\subsection{TF-Codec}

TF-Codec \cite{JiangX2023tf-codec} is equipped with various useful components, such as VQ-VAE, distance Gumbel-Softmax for rate control, and learnable 2D CNN encoder $\mathcal{E}(\cdot)$ and decoders $\mathcal{D}(\cdot)$. Moreover, as a codec that actively predicts features and conducts residual coding, it is noticeable that they tried two different architectures for prediction $\mathcal{W}^\text{pred}(\cdot)$, a 2D CNN and attention model-based one, respectively. Coupled with the custom encoder's ability to learn a suitable latent space, as well as the more sophisticated residual learner $\mathcal{W}^\ominus$ and synthesizer $\mathcal{W}^\oplus$ functions, TF-Codec achieves high-quality speech reconstruction at very low bitrates, e.g., 1 kbps.

\subsection{MDCTNet}

MDCTNet is another predictive method that works in the MDCT-transformed domain, making it compatible with the existing model-based audio codecs \cite{DavidsonG2023mdctnet}. The overall architecture is more similar to the decoder-only neural codecs, such as the WaveNet speech codec \cite{KleijnW2018wavenet} or LPCNet \cite{ValinJ2019lpcnetcoding}, than the above-mentioned predictive codecs, in the sense that prediction is not to compute the residual signal. Instead, the generative MDCTNet operates only on the decoder side to estimate the MDCT coefficients directly. 

\begin{figure}[h]
    \centering
    \includegraphics[width=0.9\linewidth]{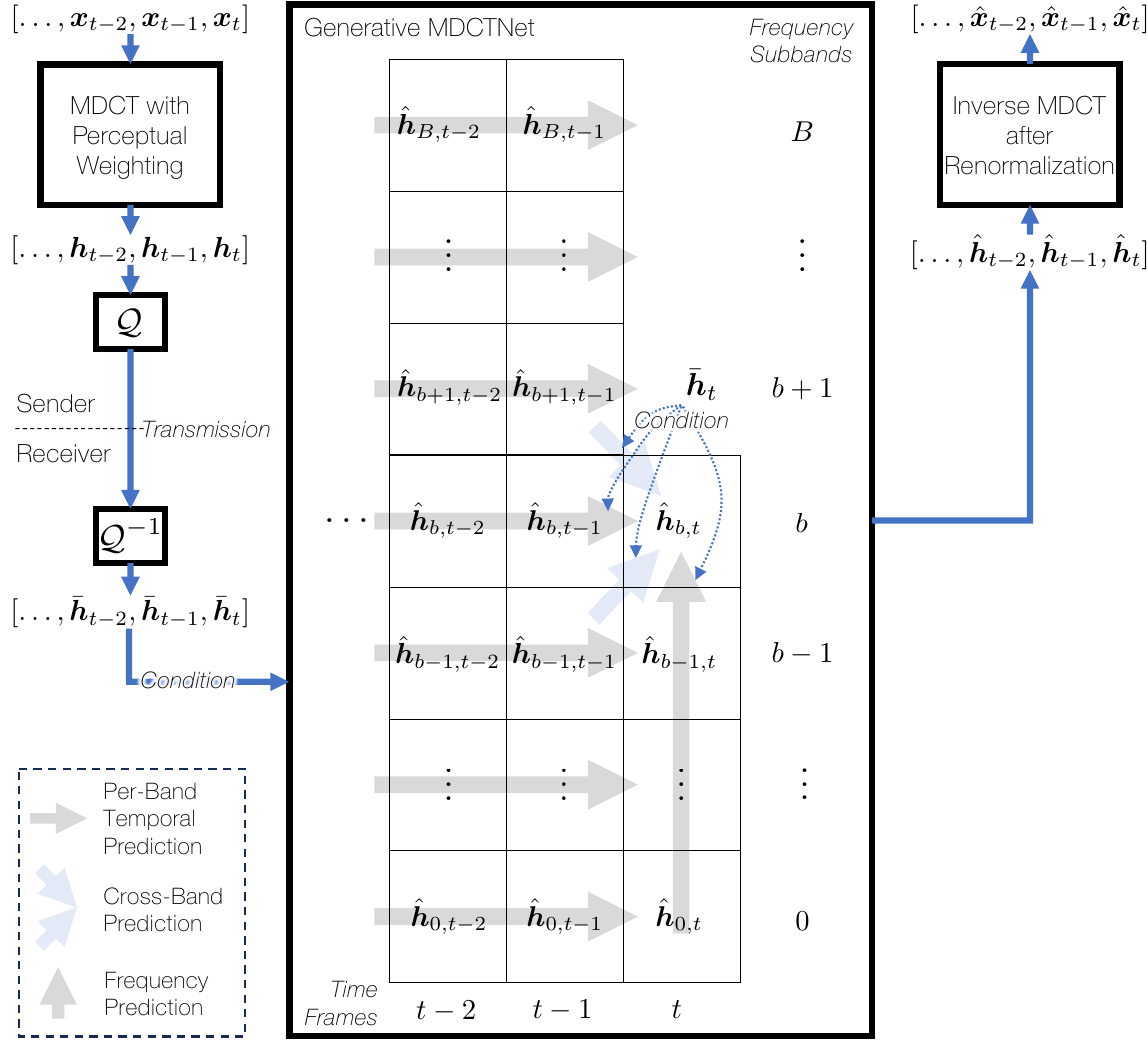}
    \caption{The simplified architecture of the MDCTNet codec. }
    \label{fig:mdctnet}
\end{figure}

Figure \ref{fig:mdctnet} illustrates the simplified MDCTNet codec architecture. On the encoder side, the input frames are transformed into a series of MDCT coefficient vectors, which are perceptually weighted via a psychoacoustic model and then quantized into the bitstream. The receiver takes the dequantized coefficient vectors $\bar{\bm h}_t$, but instead of transforming them back to the time domain directly, it only uses them to \textit{condition} the generative MDCTNet, which consists of three prediction networks. First, a temporal prediction model (a two-layer GRU module) performs per-band temporal prediction out of the past time frames,
\begin{equation}
    \hat{\bm h}_{b,t}\leftarrow\mathcal{W}_b^\text{time}(\hat{\bm h}_{b,t-1}; \bar{\bm h}_t),
\end{equation}
where $\hat{\bm h}_{b,t}$ stands for the predicted MDCT coefficient at time frame $t$ and frequency subband $f$. The GRU hidden states are omitted for notational brevity. Note that the model is conditioned with the quantized code vector $\bar{\bm h}_{b,t}$, which improves the prediction accuracy via extra information. 

In addition, MDCTNet also employs a cross-band prediction module, which is a CNN layer that takes past information from the adjacent subbands, i.e., $b-1$ and $b+1$, 
\begin{equation}
    \hat{\bm h}_{b,t}\leftarrow\mathcal{W}_b^\text{cross}(\hat{\bm h}_{b-1,t-1},\hat{\bm h}_{b+1,t-1}; \bar{\bm h}_t).
\end{equation}

Finally, the frequency-domain GRU module predicts the higher subband from the lower ones, 
\begin{equation}
    \hat{\bm h}_{b,t}\leftarrow\mathcal{W}^\text{freq}(\hat{\bm h}_{b-1,t}; \bar{\bm h}_t, \hat{\bm h}_{b,t}).
\end{equation}
where the frequency predictor is conditioned with the temporal prediction $\hat{\bm h}_{b,t}$ achieved so far. 

The MDCTNet codec also employs various other techniques that make the system more robust, such as sending the perceptual envelopes and window sequence separately, although we relegate the details to the original paper \cite{DavidsonG2023mdctnet}. In this article, we emphasize that the predictive nature of the MDCTNet mitigates the decoder's dependency on the code, $\bar{\bm h}_t$, so that its performance in the low bitrate (24 kbps, for example) is on par with Opus at 48 kbps.

\section{Psychoacoustic Models for Perceptual Loss Functions }

A fundamental issue with the data-driven approach to coding is that, during training, the quality of the decoded signals can be measured only in an objective way, e.g., using signal-to-noise ratio (SNR), etc. This is due to the nature of the gradient-based optimization algorithm that relies on the differentiation of the loss function with respect to the model parameters. Oftentimes, those objective metrics reflect only a certain aspect of the perceptual quality of the signal, leaving a gap between the perceived quality of the codec's output signal and its loss value, i.e., a signal with a high objective score does not necessarily sound good to human ears. This kind of issue can be more sophisticated if the codec has to maintain subtlety during its processing, such as for music signals with high fidelity.

One of the primary reasons behind the success of prevailing audio coding technology is its use of psychoacoustics. More specifically, according to the simultaneous masking phenomenon, a loud peak in a subband (masker) lifts up the masking threshold around it (critical band), making the nearby softer peaks less audible. Using this phenomenon, a dynamic bit allocation algorithm can adaptively assign bits based on the relationship between the quantization error and the masking threshold. 

In MPEG audio coding, for example, MP3 or advanced audio coding (AAC),  the bit allocation algorithm uses the noise-to-mask ratio (NMR). Given a masker tone's logarithmic power spectrum density (PSD) in the $b$-th critical band $ S_b$, the psychoacoustic model determines a masking threshold value $ M_f$ in all subbands $f$ affected by the masker, which defines the signal-to-mask ratio (SMR), i.e., $\text{SMR}= S_b-  M_f$. Meanwhile, if there are $m$ bits allocated for that subband $f$, which creates a noise level of $ N_f$, the signal-to-noise ratio (SNR) is defined by $\text{SNR}(m)= S_b- N_f$. Eventually, NMR is defined as the difference (in decibels) between SNR and SMR, 
\begin{equation}
    \text{NMR}(m)=\text{SMR}-\text{SNR}(m).
\end{equation}
In other words, unless the quantization noise associated with the currently available bits $m$ is louder than the masking threshold, the NMR value is negative and the noise is inaudible. Hence, the bit allocation algorithm prioritizes and assigns more bits first to the subband with the largest NMR value during its iteration.

\subsection{Psychoacoustic Calibration of Loss Functions}

Likewise, psychoacoustic models (PAM) have been the crucial component of many traditional audio coding systems. Since PAM's main usage is to make the system's behavior closer to human perception, a natural way to harmonize the concept with a data-driven method is to use it to redefine the training objectives. 

In \cite{ZhenK2020spl}, two different approaches to psychoacoustic calibration were proposed to improve the loss function's relevance to human perception. In their first proposed loss function, \textit{priority weighting}, the reconstruction loss is defined as a sum of subband-specific losses, each of which is weighted by the perceptual importance:
\begin{equation}
    \mathcal{L}_\text{PW}(t)=\sum_f w_f\big(X_f - \hat{X}_f\big)^2,
\end{equation}
where $X_f$ and $\hat{X}_f$ are the magnitude of the Fourier spectrum at subband $f$ and its reconstruction, respectively. The frame index $t$ is omitted from the equation for brevity. Since the loss is a weighted sum of subband-specific reconstruction loss, the weights $w_f$ play a big role. For a given input time domain frame $\bm x$ and its logarithmic PSD $\bm Y$, the perceptual weight vector $\bm w$ is defined by
\begin{equation}
    w_f = \log_{10} \left(\frac{10^{0.1 Y_f}}{10^{0.1 M_f}} +1\right),
\end{equation}
to give more weight to perceptually relevant subbands. For example, the weight reaches zero when the mask becomes very large while the signal is very soft, reducing the significance of the subband's contribution to the loss. The rationale behind this loss is threefold: (a) when $w_f$ is small due to the insignificant perceptual importance of $f$, the neural codec is allowed to produce a substantial amount of reconstruction error (b) with the relaxed optimization goal, a more compact neural network may still achieve the same perceptual quality (c) a lower bitrate code $\bm h$ is still acceptable as the quantization artifacts are allowed in those unimportant subbands. 

\textit{Noise modulation} is another perceptual loss derived from psychoacoustics. More similarly to MP3's bit allocation algorithm, it computes NMR and turns it into a loss function as follows:
\begin{equation}
    \mathcal{L}_\text{NM}(t)=\max_f\left(\text{ReLU}\left(\frac{N_f}{M_f}-1\right)\right).
\end{equation}
With the help from the rectified linear unit (ReLU) function, the subband with reconstruction noise $N_f$ higher than the mask $M_f$, i.e., $N_f/M_f > 1$, forms a positive loss value, although the max function follows to scan all subbands to select the most significant case. Hence, during training, an input frame is assessed by the noise modulation loss function multiple times over multiple epochs, while the most critical subband (i.e., with the highest NMR value) is addressed each time. After multiple rounds of successful optimization, the coding artifacts in the high NMR subbands are suppressed. 

In \cite{ZhenK2020spl}, it was reported that the priority weighting loss based on PAM \cite{PainterT2000ieeeproc}, when it was added to the ordinary reconstruction loss terms, led to a bitrate reduction from 79 to 64 kbps as well as a model size reduction by half while maintaining the same perceptual quality. Coupled with the noise modulation loss, the 64 kbps model achieves about 6 MUSHRA points higher than the priority weighting-only model, showcasing the merit of the combination of the two loss functions. Follow-up works investigated similar PAM-based loss functions on speech coding \cite{ByunJ2021pam}.

\section{Conclusion}

In this paper, recent neural speech and audio coding systems were presented as an example of successful harmonization of model-based and data-driven approaches. Various model-based approaches have been proposed and commercialized in the past few decades, which an entirely data-driven approach cannot easily catch up with due to the highly subjective evaluation process of the speech and audio codecs. Carefully designed hybrid systems, which also tend to benefit from a sufficiently large architecture, can be an alternative, introducing sensible coding gain to the already-saturated conventional codecs' performance. The paper first introduced a neural network-based signal enhancer as a post-processor of existing codecs. CMRL and LPCNet were another useful type of hybrid systems, where LPC was harmonized with the neural network-based end-to-end codec and vocoder, respectively. The paper also explored predictive models that work either in the custom feature space or pre-defined transform domain. Finally, we saw that psychoacoustic models can be effectively used in the data-driven training paradigms by improving the perceptual relevance of the loss function.

\singlespacing
\bibliographystyle{IEEEtran}
\bibliography{mjkim}

\begin{thebibliography}{10}
\providecommand{\url}[1]{#1}
\csname url@samestyle\endcsname
\providecommand{\newblock}{\relax}
\providecommand{\bibinfo}[2]{#2}
\providecommand{\BIBentrySTDinterwordspacing}{\spaceskip=0pt\relax}
\providecommand{\BIBentryALTinterwordstretchfactor}{4}
\providecommand{\BIBentryALTinterwordspacing}{\spaceskip=\fontdimen2\font plus
\BIBentryALTinterwordstretchfactor\fontdimen3\font minus
  \fontdimen4\font\relax}
\providecommand{\BIBforeignlanguage}[2]{{%
\expandafter\ifx\csname l@#1\endcsname\relax
\typeout{** WARNING: IEEEtran.bst: No hyphenation pattern has been}%
\typeout{** loaded for the language `#1'. Using the pattern for}%
\typeout{** the default language instead.}%
\else
\language=\csname l@#1\endcsname
\fi
#2}}
\providecommand{\BIBdecl}{\relax}
\BIBdecl

\bibitem{mp3}
{{ISO/IEC} 11172-3:1993}, ``{Coding of moving pictures and associated audio for
  digital storage media at up to about 1.5 Mbit/s},'' 1993.

\bibitem{PainterT2000ieeeproc}
T.~Painter and A.~Spanias, ``Perceptual coding of digital audio,''
  \emph{Proceedings of the IEEE}, vol.~88, no.~4, pp. 451--515, 2000.

\bibitem{sbr}
{ISO/IEC 14496-3:2001/Amd 1:2003}, ``Information technology — coding of
  audio-visual objects — part 3: Audio — amendment 1: Bandwidth
  extension,'' 2003.

\bibitem{aac-standard}
{ISO (2006) ISO/IEC 13818-7:2006}, ``{Information technology — Generic coding
  of moving pictures and associated audio information — Part 7: Advanced
  Audio Coding (AAC)},'' 2006.

\bibitem{amr-wb-standard}
{{ITU-T} {Recommendation} G.722.2}, ``{Wideband coding of speech at around 16
  kbit/s using Adaptive Multi-Rate Wideband (AMR-WB)},'' 2003.

\bibitem{usac1}
{{ISO/IEC DIS} 23003-3}, ``Information technology -- {MPEG} audio technologies
  -- part 3: Unified speech and audio coding,'' 2011.

\bibitem{usac2}
{ISO/IEC 14496-3:2009/PDAM 3}, ``Transport of unified speech and audio coding
  ({USAC}),'' 2011.

\bibitem{evs-vad}
{3GPP TS 26.451}, ``{Codec for Enhanced Voice Services (EVS); Voice Activity
  Detection (VAD)},'' 2020.

\bibitem{opus}
\BIBentryALTinterwordspacing
J.-M. Valin, K.~Vos, and T.~B. Terriberry, ``{Definition of the Opus Audio
  Codec},'' RFC 6716, Sep. 2012. [Online]. Available:
  \url{https://www.rfc-editor.org/info/rfc6716}
\BIBentrySTDinterwordspacing

\bibitem{KankanahalliS2018icassp}
S.~Kankanahalli, ``End-to-end optimized speech coding with deep neural
  networks,'' in \emph{Proc. of the IEEE International Conference on Acoustics,
  Speech, and Signal Processing (ICASSP)}, 2018.

\bibitem{AgustssonE2017softmax}
E.~Agustsson \emph{et~al.}, ``Soft-to-hard vector quantization for end-to-end
  learning compressible representations,'' in \emph{Advances in Neural
  Information Processing Systems (NIPS)}, 2017, pp. 1141--1151.

\bibitem{Zeghidour2021soundstream}
N.~Zeghidour, A.~Luebs, A.~Omran, J.~Skoglund, and M.~Tagliasacchi,
  ``Soundstream: An end-to-end neural audio codec,'' \emph{IEEE/ACM Trans.
  Audio, Speech and Lang. Proc.}, vol.~30, p. 495–507, jan 2022.

\bibitem{DefossezA2023encodec}
A.~D{\'e}fossez, J.~Copet, G.~Synnaeve, and Y.~Adi, ``High fidelity neural
  audio compression,'' \emph{Transactions on Machine Learning Research}, 2023.

\bibitem{KumarR2023dac}
R.~Kumar, P.~Seetharaman, A.~Luebs, I.~Kumar, and K.~Kumar, ``High-fidelity
  audio compression with improved {RVQGAN},'' in \emph{Advances in Neural
  Information Processing Systems (NeurIPS)}, 2023.

\bibitem{OordA2016wavenet}
A.~{van den Oord} \emph{et~al.}, ``{WaveNet: A Generative Model for Raw
  Audio},'' in \emph{Proc. 9th ISCA Workshop on Speech Synthesis Workshop (SSW
  9)}, 2016, p. 125.

\bibitem{codec2}
D.~Rowe. (2011, http://www.tapr.org/pdf/DCC2011-Codec2-VK5DGR.pdf) Codec 2-
  open source speech coding at 2400 bits/s and below [online].

\bibitem{KleijnW2018wavenet}
W.~B. Kleijn \emph{et~al.}, ``Wave{N}et based low rate speech coding,'' in
  \emph{Proc. of the IEEE International Conference on Acoustics, Speech, and
  Signal Processing (ICASSP)}, 2018, pp. 676--680.

\bibitem{GarbaceaC2019vqvae}
C.~Garbacea, A.~{van den Oord}, and Y.~Li, ``Low bit-rate speech coding with
  {VQ-VAE} and a {WaveNet} decoder,'' in \emph{Proc. of the IEEE International
  Conference on Acoustics, Speech, and Signal Processing (ICASSP)}, 2019.

\bibitem{OShaughnessyD2023speechcoding}
D.~O’Shaughnessy, ``Review of methods for coding of speech signals,''
  \emph{EURASIP Journal on Audio, Speech, and Music Processing}, vol.~8, 2023.

\bibitem{ValinJ2019lpcnetcoding}
J.-M. Valin and J.~Skoglund, ``A real-time wideband neural vocoder at 1.6 kb/s
  using {LPCNet},'' in \emph{Proc. Interspeech}, 2019.

\bibitem{BessetteB2002amrwb}
B.~Bessette \emph{et~al.}, ``The adaptive multirate wideband speech codec
  ({AMR-WB}),'' \emph{IEEE Transactions on Speech and Audio Processing},
  vol.~10, no.~8, pp. 620--636, 2002.

\bibitem{mushra}
{{ITU-R} {Recommendation} {BS} 1534-3}, ``Method for the subjective assessment
  of intermediate quality levels of coding systems ({MUSHRA}),'' 2015.

\bibitem{ValinJM2023icassp}
J.-M. Valin, J.~B\"uthe, and A.~Mustafa, ``Low-bitrate redundancy coding of
  speech using a rate-distortion-optimized variational autoencoder,'' in
  \emph{Proc. of the IEEE International Conference on Acoustics, Speech, and
  Signal Processing (ICASSP)}, 2023.

\bibitem{ZahoZ2019cnn_enhance_coded_speech}
Z.~Zhao, H.~Liu, and T.~Fingscheidt, ``Convolutional neural networks to enhance
  coded speech,'' \emph{IEEE/ACM Transactions on Audio, Speech, and Language
  Processing}, vol.~27, no.~4, pp. 663--678, 2019.

\bibitem{postefilter}
{Recommendation G.711.1}, ``{Wideband embedded extension for ITU-T G.711 pulse
  code modulation},'' 2012.

\bibitem{DengJ2019lstm-mp3-restoration}
J.~Deng \emph{et~al.}, ``{Exploiting time-frequency patterns with LSTM-RNNs for
  low-bitrate audio restoration},'' vol.~32, no.~4, pp. 1095--1107, February
  2020.

\bibitem{GoodfellowI2014gan}
I.~Goodfellow \emph{et~al.}, ``Generative adversarial nets,'' in \emph{Advances
  in neural information processing systems}, 2014, pp. 2672--2680.

\bibitem{BiswasA2020gan-coding}
A.~Biswas and D.~Jia, ``Audio codec enhancement with generative adversarial
  networks,'' in \emph{Proc. of the IEEE International Conference on Acoustics,
  Speech, and Signal Processing (ICASSP)}, 2020, pp. 356--360.

\bibitem{KorseS2022postgan}
S.~Korse, N.~Pia, K.~Gupta, and G.~Fuchs, ``{PostGAN}: A gan-based
  post-processor to enhance the quality of coded speech,'' in \emph{Proc. of
  the IEEE International Conference on Acoustics, Speech, and Signal Processing
  (ICASSP)}, 2022, pp. 831--835.

\bibitem{LC3}
{European Telecommunications Standards Institute (ETSI)}, ``{TR 103 590:
  Digital Enhanced Cordless Telecommunications (DECT); Study of Super Wideband
  Codec in DECT for narrowband, wideband and super-wideband audio communication
  including options of low delay audio connections},'' 2018.

\bibitem{FantG1971source-filter}
G.~Fant, \emph{{Acoustic theory of speech production: with calculations based
  on X-ray studies of Russian articulations}}.\hskip 1em plus 0.5em minus
  0.4em\relax Walter de Gruyter, 1971, no.~2.

\bibitem{ZhenK2019interspeech}
K.~Zhen, J.~Sung, M.~S. Lee, S.~Beack, and M.~Kim, ``Cascaded cross-module
  residual learning towards lightweight end-to-end speech coding,'' in
  \emph{Proc. Interspeech}, 2019.

\bibitem{ZhenK2022taslp}
------, ``Scalable and efficient neural speech coding: A hybrid design,''
  \emph{IEEE/ACM Transactions on Audio, Speech, and Language Processing},
  vol.~30, pp. 12--25, 2022.

\bibitem{KalchbrennerN2018wavernn}
N.~Kalchbrenner \emph{et~al.}, ``Efficient neural audio synthesis,'' in
  \emph{Proc. of the International Conference on Machine Learning (ICML)},
  vol.~80, 2018, pp. 2410--2419.

\bibitem{ChoK2014emnlp}
K.~Cho \emph{et~al.}, ``{Learning Phrase Representations using RNN
  Encoder–Decoder for Statistical Machine Translation},'' in \emph{Proc. of
  the Conference on Empirical Methods in Natural Language Processing (EMNLP)},
  2014.

\bibitem{YangH2023lpcnet}
H.~Yang, W.~Lim, and M.~Kim, ``Neural feature predictor and discriminative
  residual coding for low-bitrate speech coding,'' in \emph{Proc. of the IEEE
  International Conference on Acoustics, Speech, and Signal Processing
  (ICASSP)}, 2023.

\bibitem{JiangX2023tf-codec}
X.~Jiang, X.~Peng, H.~Xue, Y.~Zhang, and Y.~Lu, ``Latent-domain predictive
  neural speech coding,'' \emph{IEEE/ACM Transactions on Audio, Speech, and
  Language Processing}, vol.~31, pp. 2111--2123, 2023.

\bibitem{DavidsonG2023mdctnet}
G.~Davidson \emph{et~al.}, ``High quality audio coding with {MDCTNet},'' in
  \emph{Proc. of the IEEE International Conference on Acoustics, Speech, and
  Signal Processing (ICASSP)}, 2023.

\bibitem{ZhenK2020spl}
K.~Zhen, M.~S. Lee, J.~Sung, S.~Beack, and M.~Kim, ``Psychoacoustic calibration
  of loss functions for efficient end-to-end neural audio coding,'' \emph{IEEE
  Signal Processing Letters}, vol.~27, pp. 2159--2163, 2020.

\bibitem{ByunJ2021pam}
J.~Byun, S.~Shin, Y.~Park, J.~Sung, and S.~Beack, ``{Development of a
  Psychoacoustic Loss Function for the Deep Neural Network (DNN)-Based Speech
  Coder},'' in \emph{Proc. Interspeech}, 2021, pp. 1694--1698.

\end{thebibliography}
\end{document}